\documentclass[letterpaper,10pt]{ieeeconf}

\let\labelindent\relax
\usepackage{amsthm}
\usepackage{amsmath,amsfonts,amssymb,mathtools}
\usepackage{bm}
\usepackage{cases}
\usepackage[top=.75in,right=.75in,left=.75in,bottom=0.75in]{geometry}
\usepackage{color}
\usepackage[noadjust]{cite}
\usepackage[belowskip=-15pt,aboveskip=0pt]{caption}
\usepackage{subcaption}
\usepackage{arydshln}
\usepackage[inline]{enumitem}
\usepackage{url}
\usepackage{microtype}
\usepackage{multirow}
\usepackage{stfloats}
\usepackage{cuted}
\newtheorem{definition}{Definition}

\newtheorem*{notation*}{Notation}
\newtheorem{theorem}{Theorem}
\newtheorem{lemma}{Lemma}
\newtheorem{proposition}{Proposition}

\theoremstyle{definition}

\newcommand{\hop}{\mathsf{H} }
\newcommand{\rank}{\operatorname{rank} }

\IEEEoverridecommandlockouts
\title{\LARGE \bf A Frequency-Domain Version of Willems' Fundamental Lemma}
\author{T.~J.~{Meijer}$\,^{\dagger}$, S.A.N. Nouwens, V.~S.~{Dolk}, W.~P.~M.~H.~{Heemels}
\thanks{This research received funding from the European Research Council (ERC) under the Advanced ERC grant agreement PROACTHIS, no. 101055384.}%
\thanks{Tomas Meijer, Sven Nouwens and Maurice Heemels are with the Department of Mechanical Engineering, Eindhoven University of Technology, P.O. Box 513, 5600 MB Eindhoven, The Netherlands. {\tt\small \{t.j.meijer; s.a.n.nouwens; m.heemels\}@tue.nl}}%
\thanks{Victor Dolk is with ASML, De Run 6665, 5504 DT Veldhoven, The Netherlands. (e-mail: victor.dolk@asml.com)}%
\thanks{$^\dagger$ Corresponding author: T.~J.~{Meijer}.}}
\date{\today}

\begin{document}

\maketitle

\begin{abstract}
    Willems' fundamental lemma has recently received an impressive amount of attention in the (data-driven) control community. In this paper, we formulate a frequency-domain equivalent of this lemma. In doing so, we bridge the gap between recent developments in data-driven analysis and control and the extensive knowledge on non-parametric frequency-domain identification that has accumulated, particularly in industry, through decades of working with classical (frequency-domain) control and identification techniques. Our formulation also allows for the combination of multiple data sets in the sense that, in the data, multiple input directions may be excited at the same frequency. We also illustrate the usefulness of our results by demonstrating how they can be applied to perform frequency-domain-data-driven simulation.
\end{abstract}


\section{Introduction}\label{sec:introduction}
\PARstart{O}{ne} of the key strengths of classical control methodologies, such as PID control/loop shaping, lies in the fact that we can design and implement controllers based--exclusively--on easily-obtained, non-parametric and insightful frequency-domain models of the plant~\cite{Franklin2010,Skogestad2005}. This is in stark contrast with many modern control techniques, such as, e.g., model predictive control~\cite{Rawlings2020} and optimal control~\cite{Naidu2002,Skogestad2005}. While these techniques offer many benefits compared to PID control, including dealing with multiple-input multiple-output (MIMO) systems, handling constraints and guaranteed optimality/performance, they often require intricate first-principle modelling or state-space identification techniques to facilitate their design and implementation. The field of direct data-driven control aims to attain the best of both worlds by combining design and implementation using data with the many benefits of modern control techniques, see, e.g.,~\cite{DePersis2020,Coulson2019,Berberich2021}.

In recent years, Willems' fundamental lemma (WFL)~\cite{Willems2005} has become one of the cornerstones of modern data-driven control and has found numerous applications including data-driven predictive control, such as, e.g.,~\cite{Coulson2019,Berberich2021,Baros2022}, optimal control~\cite{vanWaarde2020,vanWaarde2020b,vanWaarde2023,DePersis2020}, feedback stabilization~\cite{DePersis2020}, and so on. This celebrated result states that, for a linear time-invariant (LTI) system, a single data sequence--consisting of $N$ input-output data points, where the input is persistently exciting (PE)--can be used to characterize all possible input-output solutions of length $L<N$~\cite{Berberich2020,Verhoek2021}. This idea was extended in~\cite{vanWaarde2020} to allow multiple data sequences, which are collectively persistently exciting but not necessarily individually PE, to be used to characterize all length-$L$ input-output trajectories of the system. Moreover, the desire to apply such data-driven techniques to more diverse and increasingly complicated systems has also caused considerable effort to be directed towards generalizing/adapting WFL to accomodate such systems. Examples include nonlinear systems, see, e.g.,~\cite{Berberich2020}, linear parameter-varying systems~\cite{Verhoek2021}, descriptor systems~\cite{Schmitz2022,Faulwasser2023} and stochastic systems~\cite{Faulwasser2023,Pan2022} to name a few. We refer to~\cite[Table 1]{Faulwasser2023} for a more complete overview. Interestingly, none of the aforementioned literature deals with frequency-domain data despite the fact that there often can be significant benefits to using frequency-domain data: Firstly, the use of excitation signals whose spectra are sparse in the frequency domain, e.g., multi-sine excitation, allows noise (which typically occurs at other frequencies) to be removed conveniently~\cite{Pintelon2012}. Secondly, methods, such as the local polynomial method, can be used to remove transient effects from the data, see, e.g.,~\cite{Evers2021,Pintelon2012}. Thirdly, frequency-domain methods can handle unstable plants through closed-loop identification techniques~\cite{Pintelon2012}. Last, but not least, it allows engineers to benefit from the available expertise and techniques for (non-parametric) frequency-domain identification, which has accumulated through decades of working with classical frequency-domain control/identification methodologies, particularly, in industry. 

An exception to the aforementioned focus on time-domain data is found in~\cite{Ferizbegovic2021}, where a version of WFL based on frequency-domain data is proposed. \cite{Ferizbegovic2021} characterizes all length-$L$ trajectories of the system based on samples of a single input spectrum and the corresponding cross-spectrum. Another noteworthy recent work connecting time and frequency domain in the direct data-driven setting is~\cite{Markovsky2024}, which presents a method to evaluate a frequency response directly using time-domain data without spectral leakage. While the result presented in~\cite{Ferizbegovic2021} is powerful, it faces several important limitations: \begin{enumerate*}[label=(\alph*)] \item The data-generating system is assumed to be stable, \item\label{item:single-exp} only a single input direction can be excited per frequency at which the spectra are measured, and \item the symmetry in the spectra is not appropriately accounted for. \end{enumerate*} In fact, we will see (in Section~\ref{sec:fd-wfl}) that the latter means that the formulation in~\cite{Ferizbegovic2021} is incomplete without imposing additional constraints. In many practical multiple-input multiple-output (MIMO) applications, we have access to frequency-response-function (FRF) measurements in which multiple input directions are excited at the same frequency and, as such,~\ref{item:single-exp} means that not all available data can be utilized. In this paper, we address these limitations by introducing a frequency-domain WFL, which accommodates such data sets (Section~\ref{sec:main-results}). In order to do so, we extend the notion of collective persistence of excitation to frequency-domain data sets (Section~\ref{sec:main-results}). We also show how our results can be used to perform frequency-domain-data-driven simulation based on frequency-domain data (Section~\ref{sec:applications}). Finally, conclusions are provided in Section~\ref{sec:conclusions} and all proofs can be found in the Appendix.

\noindent {\bf Notation.} We denote $\mathbb{R}=(-\infty,\infty)$, $\mathbb{Z}=\{\hdots,-2,-1,0,1,2,\hdots\}$, $\mathbb{N}=\{0,1,2,\hdots\}$, $\mathbb{N}_{[n,m]}=\{n,n+1,\hdots,m\}$ with $n,m\in\mathbb{N}$, $\mathbb{W}=[-\pi,\pi)\subset\mathbb{R}$ and $\mathbb{W}_{+} = [0,\pi)\subset\mathbb{R}$. For a complex-valued vector $v\in\mathbb{C}^{n}$, $v^\top$, $v^\hop$ and $v^*$ denote its transpose, its complex-conjugate transpose and its complex conjugate, respectively, whereas $\operatorname{Re}v$ and $\operatorname{Im}v$ denote its real and imaginary part. The notation $(u,v)$ stands for $\begin{bmatrix} u^\top & v^\top \end{bmatrix}^\top$. We denote $j=\sqrt{-1}$ and, for a matrix $A\in\mathbb{C}^{n\times m}$, $A(\star)^\hop$ is used to denote $AA^\hop$. Moreover, $e_i\in\mathbb{R}^{n}$, $i\in\mathbb{N}_{[1,n]}$, denotes the $i$-th $n$-dimensional elementary basis vector. We use boldface to denote a sequence, i.e., $\bm{x}=\{x_k\}_{k\in\mathbb{N}_{[0,N-1]}}$, $N\in\mathbb{N}$, and the notation $x_{[n,m]}=(x_n,x_{n+1},\hdots,x_m)$, $n,m\in\mathbb{N}_{[0,N-1]}$ to denote a vector containing vertically-stacked elements of a part of the sequence $\bm{x}$. Moreover, we use $\mathcal{R}^{n}_{N}\coloneqq \{\{x_k\}_{k\in\mathbb{N}_{[0,N-1]}}\mid x_k\in\mathbb{R}^{n},k\in\mathbb{N}_{[0,N-1]}\}$, $\mathcal{C}^{n}_{N}\coloneqq \{\{x_k\}_{k\in\mathbb{N}_{[0,N-1]}}\mid x_k\in\mathbb{C}^{n},k\in\mathbb{N}_{[0,N-1]}\}$ and $\mathcal{W}^+_N\coloneqq \{\{\omega_k\}_{k\in\mathbb{N}_{[0,N-1]}}\mid\omega_k\in\mathbb{W}_+,k\in\mathbb{N}_{[0,N-1]}\}$ to denote the sets of sequences of length $N$ taking values in $\mathbb{R}^{n}$, $\mathbb{C}^{n}$ and $\mathbb{W}_+$, respectively. For any $\bm{u}\in\mathcal{C}^{n}_{N}$ and $\bm{v}\in\mathcal{C}^{m}_{N}$, we denote $\{\bm{u},\bm{v}\}=\{(u_k,v_k)\}_{k\in\mathbb{N}_{[0,N-1]}}\in\mathcal{C}^{n+m}_{N}$. The Kronecker product is denoted by $\otimes$. Finally, let $H_L(\bm{x})$ for $\bm{x}\in\mathcal{R}^{n_x}_{N}$ denote the depth-$L$, with $L\leqslant N$, Hankel matrix
\begin{equation*}
    H_L(\bm{x}) = \begin{bmatrix}
        x_0 & x_1 & \hdots & x_{N-L}\\
        x_1 & x_2 & \hdots & x_{N-L+1}\\
        \vdots & \vdots & \ddots & \vdots\\
        x_{L-1} & x_{L} & \hdots & x_{N-1}
    \end{bmatrix},
\end{equation*}
induced by the sequence $\bm{x}$. 

\section{Problem statement}\label{sec:problem-statement}
Let $\Sigma$ be an LTI system, which is described by a realization $(A,B,C,D)$, i.e.,
\begin{subnumcases}{\label{eq:td-system}\Sigma~\colon~}
    x_{k+1} &= $Ax_k + Bu_k,$\label{eq:td-system-state}\\
    y_k &= $Cx_k + Du_k,$\label{eq:td-system-output}
\end{subnumcases}
with $(A,B)$ controllable, and whose transfer function $G(z)$ is given by
\begin{equation}
    G(z) = C(zI-A)^{-1}B + D,\quad z\in\mathbb{C}.
    \label{eq:tf}
\end{equation}
As mentioned before, WFL states that a single input-output trajectory of $\Sigma$, of which the input is PE, can be used to represent all of its length-$L$ input-output trajectories. For the present paper, however, the generalization of WFL to multiple data sets, which uses multiple input-output trajectories of $\Sigma$ with collectively persistently exciting input sequences to characterize all of its length-$L$ input-output trajectories~\cite{vanWaarde2020}, is particularly relevant. To formalize this extension of WFL, we introduce the following definitions:
\begin{definition}\label{dfn:trajectory}
    A pair of sequences $\{\bm{u},\bm{y}\}$ with $\bm{u}\in\mathcal{R}^{n_u}_{N}$ and $\bm{y}\in\mathcal{R}^{n_y}_{N}$ is called an input-output trajectory of $\Sigma$ in~\eqref{eq:td-system}, if there exists a state sequence $\bm{x}\in\mathcal{R}^{n_x}_{N}$ such that $\{\bm{u},\bm{x},\bm{y}\}$ satisfies~\eqref{eq:td-system-state} for $k\in\mathbb{N}_{[0,N-2]}$ and~\eqref{eq:td-system-output} for $k\in\mathbb{N}_{[0,N-1]}$.
\end{definition}
\begin{definition}[{\cite[Definition 2]{vanWaarde2020}}]\label{dfn:CPE}
    Consider $Q\in\mathbb{N}_{\geqslant 1}$ sequences $\bm{v}^i\in\mathcal{R}^{n_v}_{N}$, $i\in\mathcal{Q}\coloneqq\mathbb{N}_{[1,Q]}$. The sequences $\bm{v}^i$, $i\in\mathcal{Q}$, are called collectively persistently exciting (CPE) of order $L\in\mathbb{N}_{\geqslant 1}$, with $L\leqslant N$, if
    \begin{equation}
        \rank\begin{bmatrix}
            H_L(\bm{v}^1) & H_L(\bm{v}^2) & \hdots & H_L(\bm{v}^Q)
        \end{bmatrix}=n_vL.\label{eq:td-CPE}
    \end{equation}
\end{definition}
\noindent In the case $Q=1$, the traditional definition of PE, see~\cite[Definition 1]{vanWaarde2020}, for a single sequence is recovered from Definition~\ref{dfn:CPE}. Clearly, CPE is more flexible in the sense that it allows multiple sequences, which are not necessarily PE individually, to be combined into a data set is sufficiently ``rich''. In contrast to~\cite[Definition 2]{vanWaarde2020}, Definition~\ref{dfn:CPE} restricts the length of each individual sequence to be $N$ for the sake of simplicity (this restriction can be easily removed~\cite{vanWaarde2020}). Next, we state a version of WFL for multiple length-$N$ data sets, adapted from~\cite[Theorem 2]{vanWaarde2020}.
\begin{lemma}\label{lem:WFL}
    Let $\{\bm{u}^{d,i},\bm{y}^{d,i}\}$ with $\bm{u}^{d,i}\in\mathcal{R}^{n_u}_{N}$ and $\bm{y}^{d,i}\in\mathcal{R}^{n_y}_{N}$, $i\in\mathcal{Q}$, be $Q\in\mathbb{N}_{\geqslant 1}$ input-output trajectories of $\Sigma$. Suppose that $\bm{u}^{d,i}$, $i\in\mathcal{Q}$, are CPE of order $L+n_x$ with $L\in\mathbb{N}_{\geqslant 1}$. Then, $\{\bm{u},\bm{y}\}$ with $\bm{u}\in\mathcal{R}^{n_u}_{L}$ and $\bm{y}\in\mathcal{R}^{n_y}_{L}$ is an input-output trajectory of $\Sigma$, if and only if there exists $g\in\mathbb{R}^{Q(N-L+1)}$ such that\footnote{Here and in the sequel, the superscript $d$ denotes data collected off-line.}
    \begin{equation*}
        \begin{bmatrix}
            u_{[0,L-1]}\\
            y_{[0,L-1]}
        \end{bmatrix} = \begin{bmatrix}
            H_L(\bm{u}^{d,1}) & H_L(\bm{u}^{d,2}) & \hdots & H_L(\bm{u}^{d,Q})\\
            H_L(\bm{y}^{d,1}) & H_L(\bm{y}^{d,2}) & \hdots & H_L(\bm{y}^{d,Q})
        \end{bmatrix}g.
    \end{equation*}
\end{lemma}
\noindent We recover the standard WFL~\cite{Willems2005,vanWaarde2020} when we consider a single experiment, i.e., $Q=1$. As indicated in the introduction, most variants of WFL use time-domain data and much less effort has been directed towards the use of frequency-domain data in WFL-based data-driven control. In fact, the only frequency-domain formulation of WFL, known to us, is found in~\cite{Ferizbegovic2021}. While the results in~\cite{Ferizbegovic2021} enable the use of frequency-domain data in WFL, they only allow a single data set to be used, i.e., only one input direction may be excited at each frequency. As we will argue next, considering multiple experiments/data sets is natural, if not necessary, when dealing with frequency-domain data, e.g., FRF measurements. Indeed, when collecting FRF measurements of MIMO systems, often multiple multi-sine experiments are performed such that all input directions are excited at all frequencies~\cite{Pintelon2012}. Therefore, in many applications we have access to $M\in\mathbb{N}_{\geqslant 1}$ FRF measurements of the transfer function matrix $G(e^{j\omega})$, i.e., $\{G(e^{j\omega_m})\}_{m\in\mathcal{M}}$, $\mathcal{M}\coloneqq \mathbb{N}_{[0,M-1]}$, at the corresponding frequencies $\omega_m\in\mathbb{W}_+$, $m\in\mathcal{M}$. The formulation in~\cite{Ferizbegovic2021} is not able to exploit all of the available data, as only a single input direction per frequency can be utilized. In addition,~\cite{Ferizbegovic2021} requires the system $\Sigma$ to be stable and does not account for symmetry in the input-output spectra, which may result in complex-valued input-output trajectories (as we will see in Section~\ref{sec:fd-wfl}).

In this paper, we address the above gap in the literature by developing a frequency-domain counterpart to the multiple-experiment version of WFL in Lemma~\ref{lem:WFL}. In other words, we aim to characterize all possible length-$L$ solutions of the system $\Sigma$ using frequency-domain data, which was generated using CPE inputs. In doing so, we obtain a formulation of WFL that, unlike~\cite{Ferizbegovic2021}, accounts for symmetry and applies to unstable systems and multiple data sets. In addition, we also formulate a frequency-domain version of CPE. To underline the potential of our frequency-domain WFL, we use it to simulate an unknown system based on FRF measurements. 

\section{Main results}\label{sec:main-results}
For any (real-valued) time-domain sequence $\bm{v}=\{v_k\}_{k\in\mathbb{Z}}$ with $v_k\in\mathbb{R}^{n_v}$, $k\in\mathbb{Z}$, its spectrum $V(\omega)$, $\omega\in\mathbb{W}=[-\pi,\pi)$, is obtained by the discrete-time Fourier transform (DTFT), i.e.,
\begin{equation*}
    V(\omega) = \sum_{k=-\infty}^{\infty}v_ke^{-j\omega k},\quad \omega\in\mathbb{W}.
\end{equation*}
The time-domain sequence can be reconstructed using the inverse discrete-time Fourier transform (IDTFT), i.e.,
\begin{equation*}
    v_k = \frac{1}{2\pi}\int_{\mathbb{W}} V(\omega)e^{j\omega k}\mathrm{d}\omega,\quad k\in\mathbb{Z}.
\end{equation*}
Note that $\bm{v}$ is real-valued if and only if the corresponding spectrum $V(\omega)$ is symmetric, i.e., $V(\omega) = V^*(-\omega)$ for all $\omega\in\mathbb{W}$. As such, we deal exclusively with symmetric spectra in the sequel. Taking the DTFT of~\eqref{eq:td-system} allows us to define solutions to $\Sigma$ also in the frequency domain~\cite{Hespanha2018}.
\begin{definition}\label{dfn:fd-system}
    A pair of spectra $\{U(\omega),Y(\omega)\}$ with $U(\omega)=U^*(-\omega)\in\mathbb{C}^{n_u}$ and $Y(\omega)=Y^*(-\omega)\in\mathbb{C}^{n_y}$, $\omega\in\mathbb{W}$, is said to be an input-output spectrum of $\Sigma$, if there exists a spectrum $X(\omega)$, with $X(\omega)=X^*(-\omega)\in\mathbb{C}^{n_x}$, $\omega\in\mathbb{W}$, such that, for all $\omega\in\mathbb{W}$,
    \begin{equation}
        \begin{cases}
            e^{j\omega}X(\omega) &= AX(\omega) + BU(\omega),\\
            Y(\omega) &= CX(\omega) + DU(\omega),
        \end{cases}
        \label{eq:fd-system}
    \end{equation}
    Moreover, in that case, the triplet of spectra $\{U(\omega),X(\omega),Y(\omega)\}$ is called an input-state-output spectrum of $\Sigma$. 
\end{definition}
\noindent Clearly, $X(\omega)$ can also be eliminated from~\eqref{eq:fd-system} to find that
\begin{equation}
    Y(\omega)=G(e^{j\omega})U(\omega),\quad \omega\in\mathbb{W},
    \label{eq:X-eliminated}
\end{equation}
and, if~\eqref{eq:X-eliminated} holds, then $X(\omega)=(e^{j\omega}I-A)^{-1}BU(\omega)$, $\omega\in\mathbb{W}$, satisfies~\eqref{eq:fd-system}.

In the sequel, we will use samples of input-output spectra to characterize the dynamics of $\Sigma$ in~\eqref{eq:td-system}. Let $M\in\mathbb{N}_{\geqslant 1}$ denote the number of frequencies at which we sample the input-output spectra and let $Q\in\mathbb{N}_{\geqslant 1}$ denote the number of data sets, i.e., the number of input-output spectra being sampled. We consider $Q$ length-$M$ complex-valued sequences $\bm{U}^{d,i}\in\mathcal{C}^{n_u}_{M}$ and $\bm{Y}^{d,i}\in\mathcal{C}^{n_y}_{M}$ obtained by sampling the input-output spectra $U^{d,i}(\omega)$ and $Y^{d,i}(\omega)$, $i\in\mathcal{Q}$, respectively, at the $M$ frequencies $\bm{\omega}\in\mathcal{W}^+_M$, i.e., $U^{d,i}_m=U^{d,i}(\omega_m)\in\mathbb{C}^{n_u}$ and $Y^{d,i}_m=Y^{d,i}(\omega_m)\in\mathbb{C}^{n_y}$ for all $m\in\mathcal{M}$ and $i\in\mathcal{Q}$. Interestingly, this includes the important case described in Section~\ref{sec:problem-statement}, where we are given $M$ $n_y$-by-$n_u$ FRF measurements $G(e^{j\omega_m})$, $m\in\mathcal{M}$, of the system $\Sigma$, which are taken at the frequencies $\bm{\omega}\in\mathcal{W}^+_M$. We can incorporate this frequency-domain data by taking, for example, $U^{d,i}_m=e_i$ and $Y^{d,i}_m=G(e^{j\omega_m})U^{d,i}_m$ for $m\in\mathcal{M}$ and $i\in\mathcal{Q}$ with $Q=n_u$. In the next two sections, we will subsequently formalize the notion of (collective) persistence of excitation for such sampled spectra and use them to formulate a frequency-domain WFL.

\subsection{Frequency-domain (collective) persistence of excitation}
It should come as no surprise that the frequency-domain data, in the form of samples of the input-output spectra, needs to be sufficiently ``rich''. To formalize this notion, we define the notion of collective persistence of excitation for a sampled spectrum. To this end, for any sequence $\bm{V}\in\mathcal{C}^{n_v}_M$ and corresponding frequencies $\bm{\omega}\in\mathcal{W}^+_M$, we define $F_L(\bm{V},\bm{\omega})\in\mathbb{C}^{n_vL\times M}$ as
\begin{equation}
    F_L(\bm{V},\bm{\omega})\coloneqq \begin{bmatrix}
        W_L(\omega_0)\otimes V_0 & \hdots & W_L(\omega_{M-1})\otimes V_{M-1}
    \end{bmatrix},
\end{equation}
where $W_L(\omega)\coloneqq \begin{bmatrix} 1 & e^{j\omega} & \hdots & e^{j\omega(L-1)}\end{bmatrix}^\top$.
\begin{definition}\label{dfn:fd-CPE}
    Consider $Q\in\mathbb{N}_{\geqslant 1}$ sequences $\bm{V}^i\in\mathcal{C}^{n_v}_M$, $i\in\mathcal{Q}$, $M\in\mathbb{N}_{\geqslant 1}$, containing samples of the spectra $V^i(\omega)$, with $V^i(\omega)=(V^i(-\omega))^*$, $\omega\in\mathbb{W}$, at the frequencies\footnote{Throughout this paper, we assume for simplicity of notation that the same frequencies are sampled for each input-output spectrum. This is without loss of generality, as we can always consider different frequencies per input-output spectrum by including zeros for all other spectra at these specific frequencies. Hence, in this case $\bm{\omega}\in\mathcal{W}^{+}_M$ represents the collection of frequencies at which at least one spectrum is sampled. The resulting zero columns can be removed for the sake of computational efficiency.} $\bm{\omega}\in\mathcal{W}^{+}_M$, i.e., $V^{i}_m=V^i(\omega_m)$ for $m\in\mathcal{M}$ and $i\in\mathcal{Q}$. Then, $\bm{V}^{i}$, $i\in\mathcal{Q}$, are called CPE of order $L\in\mathbb{N}_{\geqslant 1}$, if the matrix
    \begin{equation}
        \begin{bmatrix}\begin{smallmatrix}
            F_L(\bm{V}^{1},\bm{\omega}) & \hdots & F_L(\bm{V}^{Q},\bm{\omega}) & F_L^\star(\bm{V}^{1},\bm{\omega}) & \hdots & F_L^\star(\bm{V}^{Q},\bm{\omega})
        \end{smallmatrix}\end{bmatrix}\label{eq:full-row-rank}
    \end{equation}
    is full row rank.    
\end{definition}
\noindent The following remarks are in order regarding Definition~\ref{dfn:fd-CPE}, which presents a frequency-domain version of Definition~\ref{dfn:CPE}. Firstly, observe that Definition~\ref{dfn:fd-CPE} exploits the symmetry in the spectrum by including also the complex conjugates $F_L^*(\bm{V}^{i},\bm{\omega})$, $i\in\mathcal{Q}$, in~\eqref{eq:full-row-rank}, which correspond to the spectral content at the negative frequencies. While similar rank conditions to the one in Definition~\ref{dfn:CPE} are familiar from the context of time-domain WFL-based data-driven control, see, e.g.,~\cite{vanWaarde2020,Berberich2020,Berberich2021}, we note that~\eqref{eq:full-row-rank} is equivalent to
\begin{align}
        &\operatorname{Re}\begin{bmatrix}
            F_L(\bm{V}^{1},\bm{\omega}) & \hdots & F_L(\bm{V}^{Q},\bm{\omega})
        \end{bmatrix}(\star)^\hop = \label{eq:fd-cpe}\\
        &\quad \operatorname{Re}\sum_{\mathclap{m\in\mathcal{M}}} W_{L}(\omega_m)W^\hop_L(\omega_m)\otimes \sum_{\mathclap{i\in\mathcal{Q}}} V^{i}_m(V^{i}_m)^\hop \succ 0.\nonumber
\end{align}
The equivalence between~\eqref{eq:fd-cpe} and the rank condition on~\eqref{eq:full-row-rank} is an immediate consequence of the particular structure in~\eqref{eq:full-row-rank}, which is obtained by exploiting the symmetry $V^i(\omega)=(V^i(-\omega))^*$, $\omega\in\mathbb{W}$, $i\in\mathcal{Q}$. To be precise, the equivalence follows by applying the identity $\begin{bmatrix} A & A^*\end{bmatrix}(\star)^\hop= AA^\hop + (AA^\hop)^\star = 2\operatorname{Re} AA^\hop$, for any complex matrix $A$. Interestingly, this identity yields a real-valued matrix in~\eqref{eq:fd-cpe}, which is also smaller than the one in~\eqref{eq:full-row-rank} ($n_vL$-by-$n_vL$ in~\eqref{eq:fd-cpe} as opposed to $n_vL$-by-$2n_vQM$ in~\eqref{eq:full-row-rank}). In the time domain context, a similar condition to~\eqref{eq:fd-cpe} is used to provide a quantitative PE condition~\cite{Berberich2023}, which lower-bounds the matrix in~\eqref{eq:fd-cpe} with a positive constant to quantify how close the matrix is to being rank-deficient. Secondly,~\eqref{eq:fd-cpe} implies that $2MQ\geqslant Ln_v$. To see this, observe that $\rank ww^\hop \otimes vv^\hop = \rank ww^\hop \rank vv^\hop = 1$ for any vectors $v\in\mathbb{C}^{n_v}$ and $w\in\mathbb{C}^{n_w}$. The factor $2M$ arises because any excitation at $\omega_m\neq 0$ contributes at most $2Q$ to the order of PE using the fact that $V^{i}(\omega)$ is symmetric. Excitation at $\omega_m=0$ contributes at most $Q$ to the order of PE. Lastly, when $Q=1$, we recover from~\eqref{eq:full-row-rank} a condition that is close to the one in~\cite[Theorem 2]{Ferizbegovic2021}. In fact,~\cite[Theorem 2]{Ferizbegovic2021} requires the matrix in~\eqref{eq:full-row-rank} (with $Q=1$), i.e., $F_L(\bm{V}^{1},\bm{\omega})$, without the complex-conjugate part to be full row rank and, thereby, does not account for the symmetry in $V^{1}(\omega)$.

\subsection{Frequency-domain Willems' fundamental lemma}\label{sec:fd-wfl}
\begin{figure*}[!b]
    \hrule
    \begin{align}
        \left[\begin{array}{@{}c@{}}
            u_{[0,L-1]}\\\hdashline[2pt/2pt]
            y_{[0,L-1]}
        \end{array}\right]&= \left[\begin{array}{@{}c@{}}
            \Gamma_L(\{\bm{U}^{d,i}\}_{i\in\mathcal{Q}},\bm{\omega})\\\hdashline[2pt/2pt]
            \Gamma_L(\{\bm{Y}^{d,i}\}_{i\in\mathcal{Q}},\bm{\omega})
        \end{array}\right]g\label{eq:fd-wfl}\\
        &\coloneqq \left[\begin{array}{@{}cccccc@{}}
            \operatorname{Re} F_L(\bm{U}^{d,1},\bm{\omega}) & \hdots & \operatorname{Re} F_L(\bm{U}^{d,Q},\bm{\omega}) & \operatorname{Im} F_L(\bm{U}^{d,1},\bm{\omega}) & \hdots &  \operatorname{Im} F_L(\bm{U}^{d,Q},\bm{\omega}) \\\hdashline[2pt/2pt]
            \operatorname{Re} F_L(\bm{Y}^{d,1},\bm{\omega}) & \hdots & \operatorname{Re} F_L(\bm{Y}^{d,Q},\bm{\omega}) & \operatorname{Im} F_L(\bm{Y}^{d,1},\bm{\omega}) & \hdots & \operatorname{Im} F_L(\bm{Y}^{d,Q},\bm{\omega}) 
        \end{array}\right]g \nonumber
    \end{align}
\end{figure*}

The lemma below is the frequency-domain counterpart to~\cite[Corollary 2]{Willems2005}, which states that one can render a specific matrix with samples of both state and input spectra positive definite by injecting input spectra that are CPE. 
\begin{lemma}\label{lem:technical-lemma}
    Consider the system $\Sigma$ in~\eqref{eq:td-system} and the sequences $\bm{U}^{i}\in\mathcal{C}^{n_u}_M$, $\bm{X}^{i}\in\mathcal{C}^{n_x}_M$ and $\bm{Y}^{i}\in\mathcal{C}^{n_y}_M$, $i\in\mathcal{Q}$, $M\in\mathbb{N}_{\geqslant 1}$, containing samples of the $Q\in\mathbb{N}_{\geqslant 1}$ input-state-output spectra $\{U^{i}(\omega),X^{i}(\omega),Y^{i}(\omega)\}_{i\in\mathcal{Q}}$, $\omega\in\mathbb{W}$, at the frequencies $\bm{\omega}\in\mathcal{W}^+_M$, i.e., $U^{i}_m=U^{i}(\omega_m)$, $X^{i}_m=X^{i}(\omega_m)$ and $Y^{i}_m=Y^{i}(\omega_m)$ for $m\in\mathcal{M}$ and $i\in\mathcal{Q}$. Suppose that $\bm{U}^{i}$, $i\in\mathcal{Q}$, are CPE of order $L+n_x$. Then, the matrix
    \begin{equation*}
    \begin{bmatrix}\begin{smallmatrix}
        F_1(\bm{X}^{1},\bm{\omega}) & \hdots & F_1(\bm{X}^{Q},\bm{\omega}) & F_1^\star(\bm{X}^{1},\bm{\omega}) & \hdots & F_1^\star(\bm{X}^{Q},\bm{\omega})\\
        F_L(\bm{U}^{1},\bm{\omega}) & \hdots & F_L(\bm{U}^{Q},\bm{\omega}) & F_L^\star(\bm{U}^{1},\bm{\omega}) & \hdots & F_L^\star(\bm{U}^{Q},\bm{\omega})
        \end{smallmatrix}\end{bmatrix}
\end{equation*}
    is full row rank.
\end{lemma}
\noindent The proofs of Lemma~\ref{lem:technical-lemma} and other results in the sequel are found in the Appendix. Analogous to~\eqref{eq:full-row-rank} and~\eqref{eq:fd-cpe},~\eqref{eq:Lambda-pos-def} is equivalent to
    \begin{equation}
        \Lambda=\operatorname{Re}\begin{bmatrix}
            F_1(\bm{X}^{1},\bm{\omega}) & \hdots & F_1(\bm{X}^{Q},\bm{\omega})\\
            F_L(\bm{U}^{1},\bm{\omega}) & \hdots & F_L(\bm{U}^{Q},\bm{\omega})            
            \end{bmatrix}(\star)^\hop\succ 0,
        \label{eq:Lambda-pos-def}
    \end{equation}
where similar remarks to the ones regarding~\eqref{eq:fd-cpe} and its equivalence to~\eqref{eq:full-row-rank} apply. Lemma~\ref{lem:technical-lemma} will prove useful in developing our main result in Theorem~\ref{thm:fd-wfl}, below. Moreover, the time-domain version of~\eqref{eq:Lambda-pos-def} has been shown to be useful for state feedback synthesis (when $L=1$), see, e.g.,~\cite{DePersis2020,vanWaarde2020}, or for subspace identification.

\begin{theorem}\label{thm:fd-wfl}
    Consider the system $\Sigma$ in~\eqref{eq:td-system} and the sequences $\bm{U}^{d,i}\in\mathcal{C}^{n_u}_{M}$ and $\bm{Y}^{d,i}\in\mathcal{C}^{n_y}_{M}$, $i\in\mathcal{Q}$, $M\in\mathbb{N}_{\geqslant 1}$, containing samples of the $Q\in\mathbb{N}_{\geqslant 1}$ input-output spectra $\{U^{d,i}(\omega),Y^{d,i}(\omega)\}_{i\in\mathcal{Q}}$ at the frequencies $\bm{\omega}\in\mathcal{W}^+_M$. Suppose that $\bm{U}^{d,i}$, $i\in\mathcal{Q}$, are CPE of order $L+n_x$. Then, $\{\bm{u},\bm{y}\}$, with $\bm{u}\in\mathcal{R}^{n_u}_{L}$ and $\bm{y}\in\mathcal{R}^{n_y}_{L}$, is an input-output trajectory of $\Sigma$ if and only if there exists $g\in\mathbb{R}^{2MQ}$ such that~\eqref{eq:fd-wfl} at the bottom of this page holds. 
\end{theorem}
\noindent Theorem~\ref{thm:fd-wfl} shows that, similar to Lemma~\ref{lem:WFL} based on time-domain data, we can also characterize all length-$L$ solutions of $\Sigma$ in the time domain using samples of input-output spectra that have been collected off-line. For the results in Theorem~\ref{thm:fd-wfl} to apply, the samples of the input spectra $\bm{U}^{d,i}$, $i\in\mathcal{Q}$, should be CPE. As in Lemma~\ref{lem:WFL}, samples of multiple input-output spectra, which do not necessarily satisfy this PE condition when considered separately, can be combined to achieve this. Interestingly, the time delays in $H_L(\bm{u}^{d,i})$ and $H_L(\bm{y}^{d,i})$, $i\in\mathcal{Q}$, in Lemma~\ref{lem:WFL} translate to linear phase terms, i.e., the increasing powers of $e^{j\omega_m}$ in $W_L(\omega_m)$ in~\eqref{eq:fd-wfl}, analogous to the shift theorem of the DTFT~\cite{Proakis1996}. 

For $Q=1$, the single-experiment case is recovered from~\eqref{eq:fd-wfl} resulting in a condition that is close to the one in~\cite{Ferizbegovic2021}, which states that $\{\bm{u},\bm{y}\}$, with $\bm{u}\in\mathcal{R}^{n_u}_{L}$ and $\bm{Y}\in\mathcal{R}^{n_y}_{L}$, is a trajectory of $\Sigma$ if and only if there exists $G\in\mathbb{C}^{M}$ such that
\begin{equation}
    \begin{bmatrix}
        u_{[0,L-1]}\\
        y_{[0,L-1]}
    \end{bmatrix} = \begin{bmatrix}
        F_L(\bm{U}^{d,1},\bm{\omega})\\
        F_L(\bm{Y}^{d,1},\bm{\omega})
    \end{bmatrix}G.\label{eq:fd-WFL-schon}
\end{equation}
In addition,~\cite{Ferizbegovic2021} requires the underlying system $\Sigma$ to be stable, which is not necessary for Theorem~\ref{thm:fd-wfl}. In fact, Section~\ref{sec:applications} presents an application of Theorem~\ref{thm:fd-wfl} to perform data-driven simulation of an \emph{unstable} plant. We also observe that~\cite{Ferizbegovic2021} does not necessarily account for the symmetry in the input-output spectra $\{U^{d,i}(\omega),Y^{d,i}(\omega)\}_{i\in\mathcal{Q}}$ since~\eqref{eq:fd-WFL-schon} does not necessarily contain the conjugate data $F_L^*(\bm{U}^{d,1},\bm{\omega})$ and $F_L^*(\bm{Y}^{d,1},\bm{\omega})$. This not only results in a lower order of PE compared to Theorem~\ref{thm:fd-wfl} using the same frequency-domain data but also means that~\eqref{eq:fd-WFL-schon} is incomplete without additional constraints on $G$. To see the latter, take any $G\in\mathbb{R}^{M}\setminus\{0\}$, then we find that $\bm{u}$ and $\bm{y}$ are complex-valued. Finally, by exploiting symmetry, we have formulated~\eqref{eq:fd-wfl} as a real-valued matrix-vector product, whereas~\eqref{eq:fd-WFL-schon} involves a complex-valued matrix-vector product.

\section{Applications}\label{sec:applications}
\subsection{FRF-based data-driven simulation}
In this section, we illustrate how our results can be used to simulate the unknown system $\Sigma$ using input-output spectra that have been collected off-line. Since no state-space realization of $\Sigma$ is available, the initial condition $x_0$ is imposed through an initial input-output trajectory of at least length $n_x$. Similar data-driven simulation algorithms have been proposed in the time domain, see, e.g.,~\cite{Markovsky2008,Berberich2020,Ferizbegovic2021}.
\begin{proposition}\label{prop:data-driven-simulation}
    Consider the system $\Sigma$ in~\eqref{eq:td-system} and the sequences $\bm{U}^{d,i}\in\mathcal{C}^{n_u}_M$ and $\bm{Y}^{d,i}\in\mathcal{C}^{n_y}_{M}$, $i\in\mathcal{Q}$, $M\in\mathbb{N}_{\geqslant 1}$ containing samples of the $Q\in\mathbb{N}_{\geqslant 1}$ input-output spectra $\{U^{d,i}(\omega),Y^{d,i}(\omega)\}_{i\in\mathcal{Q}}$ at the frequencies $\bm{\omega}\in\mathcal{W}^{+}_M$. Let $L_0\in\mathbb{N}_{\geqslant n_x}$ and suppose that $\bm{U}^{d,i}(\omega)$, $i\in\mathcal{Q}$, are CPE of order $L+n_x$ with $L\geqslant L_0$. Let $\{\bm{u}^{\mathrm{ini}},\bm{y}^{\mathrm{ini}}\}$, with $\bm{u}^{\mathrm{ini}}\in\mathcal{R}^{n_u}_{L_0}$ and $\bm{y}^{\mathrm{ini}}\in\mathcal{R}^{n_y}_{L_0}$, be an (initial) input-output trajectory of length $L_0$. If $\{\bm{u},\bm{y}\}$, with $\bm{u}\in\mathcal{R}^{n_u}_L$, $\bm{y}\in\mathcal{R}^{n_y}_L$ and $\{u_k,y_k\}_{k\in\mathbb{N}_{[0,L_0-1]}}=\{\bm{u}^{\mathrm{ini}},\bm{y}^{\mathrm{ini}}\}$, is an input-output trajectory of $\Sigma$, then there exists $g\in\mathbb{R}^{2MQ}$ such that
    \begin{equation}
        \left[\begin{array}{@{}c@{}}
            u^{\mathrm{ini}}_{[0,L_0-1]}\\
            u_{[L_0,L-1]}\\\hdashline[2pt/2pt]
            y^{\mathrm{ini}}_{[0,L_0-1]}
        \end{array}\right] = \left[\begin{array}{@{}c@{}}
            \Gamma_L(\{\bm{U}^{d,i}\}_{i\in\mathcal{Q}},\bm{\omega})\\            \hdashline[2pt/2pt]
            \Gamma_{L_0}(\{\bm{Y}^{d,i}\}_{i\in\mathcal{Q}},\bm{\omega})
        \end{array}\right]g,\label{eq:ini}
    \end{equation}
    with $\Gamma$ as given in~\eqref{eq:fd-wfl}, holds. Moreover, in that case, the output trajectory $\bm{y}$ satisfies
    \begin{equation}
        y_{[0,L-1]} = \Gamma_{L}(\{\bm{Y}^{d,i}\}_{i\in\mathcal{Q}},\bm{\omega})g.\label{eq:future}
    \end{equation}
\end{proposition}
\noindent We can use Proposition~\ref{prop:data-driven-simulation} to perform FRF-based simulation of an unknown system by first solving~\eqref{eq:ini} for $g$ and, then, computing the future outputs $y_{[L_0,L-1]}$ using the last $(L-L_0)n_y$ rows of~\eqref{eq:future}. The correct initial condition is enforced through the initial trajectory $\{\bm{u}^{\mathrm{ini}},\bm{y}^{\mathrm{ini}}\}$. We further illustrate this using the numerical example below.

\subsection{Example}
Consider the unstable batch reactor system~\cite{Walsh2001,vanWaarde2020}, which we discretize using a sampling time of $0.5$s to obtain a system of the form~\eqref{eq:td-system} with
\begin{align}
    A &= \begin{bmatrix}
        2.622 & 0.320 & 1.834 & -1.066\\
        -0.238 & 0.187 & -0.136 & 0.202\\
        0.161 & 0.789 & 0.286 & 0.606\\
        -0.104 & 0.764 & 0.089 & 0.736
    \end{bmatrix},\label{eq:AB}\\
    B &= \begin{bmatrix}
        0.465 & -1.550\\
        1.314 & 0.085\\
        2.055 & -0.673\\
        2.023 & -0.160
    \end{bmatrix}\text{ and }C = \begin{bmatrix} 
        1 & 0 & 1 & -1\\
        0 & 1 & 0 & 0
    \end{bmatrix}.\nonumber
\end{align}
To generate the off-line data, we excite $M=10$ frequencies $\omega_m=0.1(m+1)$, $m\in\mathcal{M}$, by setting $U^{d,i}_m = e_i$, $m\in\mathcal{M}$, $i\in\mathcal{Q}\coloneqq\mathbb{N}_{[1,Q]}$ with $Q=n_u$, and computing the corresponding $Y^{d,i}_m=G(e^{j\omega_m})U^{d,i}_m$, $i\in\mathcal{Q}$, $m\in\mathcal{M}$. We verify using~\eqref{eq:fd-cpe} that the input spectra $U^{d,i}(\omega)$, $i\in\mathcal{Q}$, are CPE of order $8$. In practice, we would obviously not be able to use the model of $\Sigma$ to generate the data, as we have done here, but, instead, closed-loop FRF measurements could be collected using some pre-stabilizing controller~\cite{Pintelon2012}.

\begin{figure}[!ht]
    \centering
    \includegraphics[width=.5\textwidth,trim=0 150 0 14]{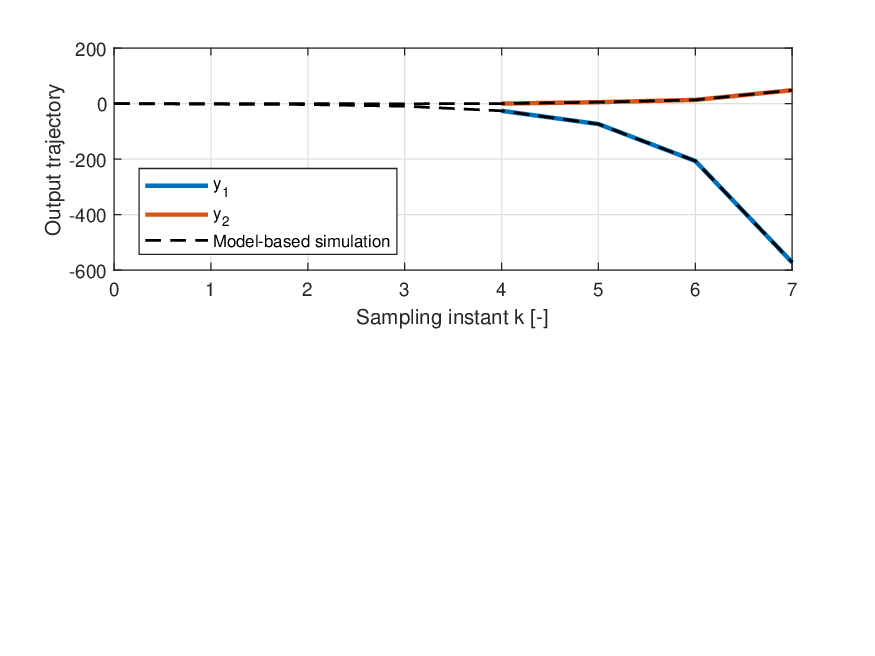}
    \caption{Data-driven simulation of an unstable batch reactor.}
    \label{fig:data-driven-simulation}
\end{figure}

To perform the simulation, we first generate an initial trajectory of length $L_0=n_x=4$ starting from the initial condition $x_0=0$ and using a randomly generated input sequence $\bm{u}\in\mathcal{R}^{n_u}_{L_0}$. Afterwards, we perform a simulation of length $L=4$ using Proposition~\ref{prop:data-driven-simulation} with a randomly generated input sequence $\{u_k\}_{k\in\mathbb{N}_{[L_0,L-1]}}$. The results are depicted in Fig.~\ref{fig:data-driven-simulation} along with the trajectory $\bm{y}^{true}\in\mathcal{C}^{n_y}_{L}$ computed by a model-based simulation (of which the first $L_0$ were used as the initial input-output trajectory $\{\bm{u}^{\mathrm{ini}},\bm{y}^{\mathrm{ini}}\}$). We see that, despite the system being unstable and the output trajectories rapidly diverging, the data-driven simulation accurately resembles the model-based simulation. In fact, we find that $\|y_{[L_0,L-1]}-y^{\mathrm{true}}_{[L_0,L-1]}\|=1.010\cdot 10^{-9}$ and $\|y_{[L_0,L-1]}-y^{\mathrm{true}}_{[L_0,L-1]}\|/\|y^{\mathrm{true}}\|=1.640\cdot 10^{-12}$.

\section{Conclusions}\label{sec:conclusions}
In this paper, we presented a frequency-domain equivalent to Willems' fundamental lemma, which can be used to fully characterize the input-output behaviour of an unknown system based purely on frequency-domain data collected off-line. Unlike existing results, we are able to make use of multiple frequency-domain data sets in the sense that, in the off-line data, multiple input directions may be excited at the same frequency. By doing so, we are able to use all of the available data, for instance, in the particularly relevant case when we have access to FRF measurements of the plant. To facilitate this, we extend the notion of collective persistence of excitation to the frequency domain. Finally, we demonstrate our results by using them to perform frequency-domain-data-driven simulation. Since the results on data-driven simulation can be used to predict future behaviour of a system, they can be naturally integrated into a frequency-domain-data-driven predictive control scheme, which is the focus of future work.

\bibliographystyle{IEEEtran}
\bibliography{phd-bibtex}

\section*{Appendix - Proofs}
\subsection{Proof of Lemma~\ref{lem:technical-lemma}}
To prove Lemma~\ref{lem:technical-lemma}, we follow similar reasoning as in~\cite{vanWaarde2020}. By construction, it is sufficient to show that $\Lambda w=0$, for any $w\in\mathbb{R}^{n_x+Ln_u}$, implies that $w=0$. To this end, let $w=(\xi,\eta)$, for some $\xi\in\mathbb{R}^{n_x}$ and $\eta\in\mathbb{R}^{n_uL}$, and suppose that $\Lambda w=0$. Let
\begin{equation*}
    \Theta =\operatorname{Re}\begin{bmatrix}
            F_1(\bm{X}^{1},\bm{\omega}) & \hdots & F_1(\bm{X}^{Q},\bm{\omega})\\
            F_{L+n_x}(\bm{U}^{1},\bm{\omega}) & \hdots & F_{L+n_x}(\bm{U}^{Q},\bm{\omega})            \end{bmatrix}(\star)^\hop.
\end{equation*}
By construction, it holds that $\Theta w_0=0$ with $w_0=(\xi,\eta,0_{n_xn_u})$. Since $w_0$ is real, it follows that
\begin{equation*}
    \operatorname{Re}\sum_{m\in\mathcal{M}}\sum_{i\in\mathcal{Q}}\left(\begin{bmatrix}
        X^{i}_m\\
        W_{L+n_x}(\omega_m)\otimes U^{i}_m
    \end{bmatrix}(\star)^\hop\begin{bmatrix}
        \xi\\
        \eta\\
        0_{n_xn_u}
    \end{bmatrix}\right)=0.
\end{equation*}
By multiplying each element in this summation with increasing powers of $e^{-j\omega_m}$, it follows that, for any $s\in\mathbb{N}_{[0,n_x]}$, 
\begin{align*}
    0&=\operatorname{Re}\sum_{{m\in\mathcal{M}}}\sum_{{i\in\mathcal{Q}}}\begin{bmatrix}
        X^{i}_m\\
        W_{L+n_x}(\omega_m)\otimes U^{i}_m
    \end{bmatrix}(\star)^\hop\begin{bmatrix}\begin{smallmatrix}
        e^{-j\omega_ms}\xi\\
        e^{-j\omega_ms}\eta\\
        0_{n_xn_u}
    \end{smallmatrix}\end{bmatrix},\\
    &=\operatorname{Re}\sum_{{m\in\mathcal{M}}}\sum_{{i\in\mathcal{Q}}}\begin{bmatrix}
        X^{i}_m\\
        W_{L+n_x}(\omega_m)\otimes U^{i}_m
    \end{bmatrix}(\star)^\hop\begin{bmatrix}\begin{smallmatrix}
        e^{-j\omega_ms}\xi\\
        0_{sn_u}\\
        \eta\\
        0_{(n_x-s)n_u}
    \end{smallmatrix}\end{bmatrix},\\
    &\stackrel{\eqref{eq:fd-system}}{=} \Theta w_s,
\end{align*}
with $w_s \coloneqq ((A^\top)^s\xi,B^\top (A^\top)^{s-1}\xi,\hdots,B^\top\xi,\eta,0_{(n_x-s)n_u})$, $s\in\mathbb{N}_{[0,n_x]}$. By the CPE assumption, the bottom-right $(L+n_x)n_u$-by-$(L+n_x)n_u$ block of $\Theta$ is positive definite (see~\eqref{eq:fd-cpe}) and, thereby, $\dim\ker \Theta\leqslant n_x$. As such, the $n_x+1$ vectors $w_s$, $s\in\mathbb{N}_{[0,n_x]}$, which we computed before, must be linearly dependent. Since the last $n_u$ entries of $w_0,w_1,\hdots,w_{n_x-1}$ are zero, the last $n_u$ entries of $\eta$ must, by inspection of $w_{n_x}$, also be zero. Hence, the last $2n_u$ entries of $w_0,w_1,\hdots,w_{n_x-1}$ are zero and, by inspection of $w_{n_x}$, the last $2n_u$ entries of $\eta$ must be zero. Repeating this process $L$ times yields $\eta=0$. 

Using the Cayley-Hamilton theorem, we have 
\begin{equation*}
    w_{n_x} + \sum_{\mathclap{i\in\mathbb{N}_{[0,n_x-1]}}} c_iw_i = (0_{n_x},\bar{w}_0,0_{n_uL}),
\end{equation*}
for some $c_i\in\mathbb{R}$, $i\in\mathbb{N}_{[0,n_x-1]}$, with
\begin{equation*}
    \bar{w}_0 = \begin{bmatrix}
        B^\top(c_1I+\hdots+c_{n_x-1}(A^\top)^{n_x-2}+(A^\top)^{n_x-1})\\
        B^\top(c_2I+\hdots+c_{n_x-1}(A^\top)^{n_x-3}+(A^\top)^{n_x-2})\\
        \vdots\\
        B^\top (c_{n_x-1}I + A)\\
        B^\top
    \end{bmatrix}\xi.
\end{equation*}
By construction, $\bar{w}\coloneqq w_{n_x}+\sum_{i\in\mathbb{N}_{[0,n_x-1]}}c_iw_i=(0_{n_x},\bar{w}_0,0_{n_uL})$ satisfies $\bar{w}\in\ker\Theta$. Thus, we conclude, due to CPE, that $\bar{w}_0=0$. Using the fact that $\Sigma$ is controllable, it follows that $\xi=0$ and, thus,~\eqref{eq:Lambda-pos-def} holds.\qed
    
\subsection{Proof of Theorem~\ref{thm:fd-wfl}}
Let $X^{d,i}(\omega)=(e^{j\omega}I-A)^{-1}BU^{d,i}(\omega)$ and let $\bm{X}^{d,i}\in\mathcal{C}^{n_x}_{M}$ be such that $X^{d,i}_m=X^{d,i}(\omega_m)$ for $m\in\mathcal{M}$, $i\in\mathcal{Q}$. Then, $\{U^{d,i}(\omega),X^{d,i}(\omega),Y^{d,i}(\omega)\}_{i\in\mathcal{Q}}$ with $X^{d,i}(\omega)$, $i\in\mathcal{Q}$, are input-state-output spectra of $\Sigma$. Since $\bm{U}^{d,i}$, $i\in\mathcal{Q}$, are CPE of order $L+n_x$, Lemma~\ref{lem:technical-lemma} implies that~\eqref{eq:Lambda-pos-def} holds.

By~\eqref{eq:td-system}, any length-$L$, $L\in\mathbb{N}_{\geqslant 1}$, input-output trajectory $\{\bm{u},\bm{y}\}$ of $\Sigma$ with $\bm{u}\in\mathcal{R}^{n_u}_{L}$ and $\bm{y}\in\mathcal{R}^{n_y}_{L}$ satisfies
\begin{equation}
    \begin{bmatrix}
        u_{[0,L-1]}\\
        y_{[0,L-1]}
    \end{bmatrix} = \begin{bmatrix}
        0 & I\\
        \mathcal{O}_L & \mathcal{T}_L
    \end{bmatrix}\begin{bmatrix}
        x_0\\
        u_{[0,L-1]}
    \end{bmatrix},
    \label{eq:uy-x0u}
\end{equation}
where
\begin{equation}
    \mathcal{O}_L=\begin{bmatrix}
        C\\
        CA\\
        \vdots\\
        CA^{L-1}
    \end{bmatrix}\hspace*{-1mm},\mathcal{T}_L=\begin{bmatrix}
        D & 0 & \hdots & 0\\
        CB & D & \hdots & 0\\
        \vdots & \vdots & \ddots & \vdots\\
        CA^{L-2}B & CA^{L-3}B & \hdots & D
    \end{bmatrix}\hspace*{-1mm}.\label{eq:OLTL}
\end{equation}
Due to~\eqref{eq:Lambda-pos-def}, for any $x_0\in\mathbb{R}^{n_x}$ and $u_{[0,L-1]}\in\mathbb{R}^{n_uL}$, there exists $g\in\mathbb{R}^{n_uL+n_x}$ such that
\begin{equation}
    \begin{bmatrix}
        x_0\\
        u_{[0,L-1]}
    \end{bmatrix} = \Lambda g.
    \label{eq:Lambda-g}
\end{equation}
Observe that $Y^{d,i}_m$, $X^{d,i}_m$ and $U^{d,i}_m$ satisfy, for $i\in\mathcal{Q}$, $m\in\mathcal{M}$,
\begin{equation}
    W_L(\omega_m)\otimes Y^{d,i}_m = \mathcal{O}_LX_m^{d,i} + \mathcal{T}_L (W_L(\omega_m)\otimes U^{d,i}_m),
    \label{eq:YXU-relation}
\end{equation}
and, thus, by substitution of~\eqref{eq:Lambda-g} and~\eqref{eq:YXU-relation} into~\eqref{eq:uy-x0u}, we obtain
\begin{align}
    &\begin{bmatrix}
        u_{[0,L-1]}\\
        y_{[0,L-1]}
    \end{bmatrix} = \begin{bmatrix}
        0 & I\\
        \mathcal{O}_L & \mathcal{T}_L
    \end{bmatrix}\Lambda g,\label{eq:in-terms-of-g}\\
    &~= \begin{bmatrix}
        0 & I\\
        \mathcal{O}_L & \mathcal{T}_L
    \end{bmatrix}\operatorname{Re}\sum_{m\in\mathcal{M}}\sum_{i\in\mathcal{Q}}\begin{bmatrix}
        X^{d,i}_{m}\\
        W_L(\omega_m)\otimes U^{d,i}_m
    \end{bmatrix}(\star)^\hop g,\nonumber\\
    &~\stackrel{\eqref{eq:YXU-relation}}{=} \operatorname{Re}\sum_{m\in\mathcal{M}}\sum_{i\in\mathcal{Q}}\begin{bmatrix}
        W_L(\omega_m)\otimes U^{d,i}_m\\
        W_L(\omega_m)\otimes Y^{d,i}_m
    \end{bmatrix}\begin{bmatrix}
        X^{d,i}_m\\
        W_L(\omega_m)\otimes U^{d,i}_m
    \end{bmatrix}^\hop g.\nonumber
\end{align}
There exists $g\in\mathbb{R}^{n_uL+n_x}$ for which~\eqref{eq:in-terms-of-g} holds if and only if there exist $G^i_m\in\mathbb{C}$, $i\in\mathcal{Q}$, $m\in\mathcal{M}$, such that
\begin{equation}
    \begin{bmatrix}
        u_{[0,L-1]}\\
        y_{[0,L-1]}
    \end{bmatrix} = \operatorname{Re}\sum_{m\in\mathcal{M}}\sum_{i\in\mathcal{Q}}\begin{bmatrix}
        W_L(\omega_m)\otimes U^{d,i}_m\\
        W_L(\omega_m)\otimes Y^{d,i}_m
    \end{bmatrix}G^i_m.\label{eq:in-terms-of-G}
\end{equation}
To see this, note that if $g\in\mathbb{R}^{n_uL+n_x}$ satisfies~\eqref{eq:in-terms-of-g}, then 
\begin{equation}
    G^i_m = \begin{bmatrix} (X^{d,i}_m)^\hop & W_L^\hop(\omega_m)\otimes (U^{d,i}_m)^\hop\end{bmatrix}g
    \label{eq:g-to-G}
\end{equation}
for $i\in\mathcal{Q}$, $m\in\mathcal{M}$ satisfy~\eqref{eq:in-terms-of-G}. To see the converse, note that, for any $G^{i}_m\in\mathbb{C}$, $i\in\mathcal{Q}$, $m\in\mathcal{M}$, we can find $g\in\mathbb{R}^{n_uL+n_x}$ such that~\eqref{eq:g-to-G} except if $X^{d,i}_m=0$ and $U^{d,i}_m=0$. However, if $X^{d,i}_m=0$ and $U^{d,i}_m=0$, then $Y^{d,i}_m=0$ such that $G^i_m$ is multiplied by zero and, hence, its value is free. Thus, the existence of $g\in\mathbb{R}^{n_uL+n_x}$ such that~\eqref{eq:in-terms-of-g} holds is equivalent to the existence of $G^{i}_m\in\mathbb{C}$, $i\in\mathcal{Q}$, $m\in\mathcal{M}$, such that~\eqref{eq:in-terms-of-G} holds. Let $\bar{G}^i=(G^i_0,G^i_1,\hdots,G^i_{M-1})\in\mathbb{C}^{M}$, $i\in\mathcal{Q}$, and $\bar{G}=(\bar{G}^1,\bar{G}^2,\hdots,\bar{G}^Q)\in\mathbb{C}^{MQ}$. Then, we can write~\eqref{eq:in-terms-of-G} as
\begin{equation*}
    \begin{bmatrix}
        u_{[0,L-1]}\\
        y_{[0,L-1]}
    \end{bmatrix} = \operatorname{Re} \begin{bmatrix}
        F_L(\bm{U}^{d,1},\bm{\omega}) & \hdots & F_L(\bm{U}^{d,Q},\bm{\omega})\\
        F_L(\bm{Y}^{d,1},\bm{\omega}) & \hdots & F_L(\bm{Y}^{d,Q},\bm{\omega})
    \end{bmatrix}\bar{G}.
\end{equation*}
Since $\operatorname{Re}(Uz)=\operatorname{Re}(U)\operatorname{Re}(z)-\operatorname{Im}(U)\operatorname{Im}(z)$, $U\in\mathbb{C}^{n\times m}$, $z\in\mathbb{C}^{m}$, we find~\eqref{eq:fd-wfl} with $g=(\operatorname{Re}\bar{G},-\operatorname{Im}\bar{G})$.\qed

\subsection{Proof of Proposition~\ref{prop:data-driven-simulation}}
Let $X^{d,i}(\omega)=(e^{j\omega}I-A)^{-1}BU^{d,i}(\omega)$ and let $\bm{X}^{d,i}\in\mathcal{C}^{n_x}_{M}$ be such that $X^{d,i}_m=X^{d,i}(\omega_m)$ for $m\in\mathcal{M}$, $i\in\mathcal{Q}$. Then, $\{U^{d,i}(\omega),X^{d,i}(\omega),Y^{d,i}(\omega)\}_{i\in\mathcal{Q}}$ with $X^{d,i}(\omega)$, $i\in\mathcal{Q}$, are input-state-output spectra of $\Sigma$. Thus, $Y^{d,i}_m$, $X^{d,i}_m$ and $U^{d,i}_m$ satisfy~\eqref{eq:YXU-relation}. Since $\bm{U}^{d,i}$, $i\in\mathcal{Q}$, are CPE of order $L+n_x$, Lemma~\ref{lem:technical-lemma} implies that~\eqref{eq:Lambda-pos-def} holds. By~\eqref{eq:td-system}, $\{\bm{u},\bm{y}\}$ satisfies
\begin{equation}
    y_{[0,L_0-1]} = \mathcal{O}_{L_0}x_0 + \mathcal{T}_{L_0}u_{[0,L_0-1]},
\end{equation}
with $\mathcal{O}_L$ and $\mathcal{T}_{L}$ as in~\eqref{eq:OLTL}. Hence, for any $u_{[0,L_0-1]}$, $y_{[0,L_0-1]}$ and $u_{[L_0,L]}$, there exists $g\in\mathbb{R}^{n_uL+n_x}$ such that
\begin{align}
    &\left[\begin{array}{@{}c@{}}
        u^{\mathrm{ini}}_{[0,L_0-1]}\\
        u_{[L_0,L]}\\\hdashline[2pt/2pt]
        y^{\mathrm{ini}}_{[0,L_0-1]}
    \end{array}\right] = \left[\begin{array}{@{}cc@{}}
        0 & I\\\hdashline[2pt/2pt]
        \mathcal{O}_{L_0} & \mathcal{T}_{L_0}
    \end{array}\right]\left[\begin{array}{@{}c@{}}
        x_0\\
        u_{[0,L-1]}
    \end{array}\right],\\
    &~\stackrel{\eqref{eq:Lambda-pos-def}}{=} \left[\begin{array}{@{}cc@{}}
        0 & I\\\hdashline[2pt/2pt]
        \mathcal{O}_{L_0} & \mathcal{T}_{L_0}
    \end{array}\right]\operatorname{Re}\sum_{m\in\mathcal{M}}\sum_{i\in\mathcal{Q}}\begin{bmatrix}
        X^{d,i}_{m}\\
        W_L(\omega_m)\otimes U^{d,i}_m
    \end{bmatrix}(\star)^\hop g,\nonumber\\
    &~\stackrel{\eqref{eq:YXU-relation}}{=} \operatorname{Re}\sum_{m\in\mathcal{M}}\sum_{i\in\mathcal{Q}}\begin{bmatrix}
        W_L(\omega_m)\otimes U^{d,i}_m\\
        W_L(\omega_m)\otimes Y^{d,i}_m
    \end{bmatrix}\begin{bmatrix}
        X^{d,i}_m\\
        W_L(\omega_m)\otimes U^{d,i}_m
    \end{bmatrix}^\hop g.\nonumber
\end{align}
The proof for~\eqref{eq:ini} is completed using the same steps as at the end of the proof of Theorem~\ref{thm:fd-wfl}. It follows immediately from Theorem~\ref{thm:fd-wfl} that $y_{[L_0,L-1]}$ is given by~\eqref{eq:future}. \qed

\end{document}